# A Comprehensive Framework for Automated Segmentation of Perivascular Spaces in Brain MRI with the nnU-Net

William Pham[a,*], Alexander Jarema[b], Donggyu Rim[a], Zhibin Chen[a], Mohamed Salah Khlif[a], Vaughan G. Macefield[a,d], Luke A. Henderson[c], Amy Brodtmann[a]

[a] Department of Neuroscience, School of Translational Medicine, Faculty of Medicine, Nursing and Health Sciences, Monash University, Australia.
[b] Department of Radiology, Alfred Health, Melbourne, Australia.
[c] Department of Anatomy and Histology, University of Sydney, Sydney, NSW, Australia.
[d] School of Medicine, Western Sydney University, Sydney, Australia.

*** Corresponding author**
William Pham
Monash University, School of Translational Medicine
Department of Neuroscience
Level 6, 75 Commercial Road,
Melbourne, VIC, 3004
Email: william.pham@monash.edu
ORCID: 0000-0003-2865-7250

**Support:** This research was supported by an Australian Government Research Training Program (RTP) Scholarship.

**Abstract**

*Background:* Enlargement of perivascular spaces (PVS) is common in neurodegenerative disorders including cerebral small vessel disease, Alzheimer's disease, and Parkinson's disease, potentially associated with impaired clearance pathways. There is a need for reliable PVS estimation methods. *Aim:* To optimise a widely used deep learning model, the no-new-UNet (nnU-Net), for PVS segmentation. *Methods:* In 30 healthy participants (mean±SD age: 50±18.9 years; 13 women), T1-weighted MRI images were acquired using three different MRI scanners (3T Siemens Tim Trio, 3T Philips Achieva, and 7T Siemens Magnetom). After standard preprocessing, PVS were manually segmented across ten axial slices in each participant. The nnU-Net was then used to analyse the remaining unlabelled slices via a sparse annotation strategy. These were manually corrected and comprised the training and validation dataset. Model performance was evaluated using 5-fold cross validation (5FCV). The main performance metric was the Sørensen-Dice Similarity Coefficient (DSC). *Results:* The voxel-spacing agnostic model (mean±SD DSC=64.3±3.3%) outperformed models which resampled images to a common resolution (DSC=40.5–55%). Model performance improved substantially following iterative label cleaning (DSC=85.7±1.2%). Semi-supervised learning with pseudo-labels (n=12,740) from 18 additional datasets improved the agreement between actual and predicted PVS cluster counts (Lin's concordance correlation coefficient=0.89, 95%CI: 0.82–0.93). The final model demonstrated robust performance across 3T and 7T MRI scans, achieving DSC of 88.5±2.4% and 86.2±2.9%, respectively. Moreover, the model capabilities were extended, allowing us to detect PVS in the midbrain (DSC=64.3±6.5%) and hippocampus (DSC=67.8±5%). *Conclusions:* Our deep learning models provide a robust and comprehensive framework for the automated quantification of PVS in brain MRI.

## 1. Introduction

Perivascular spaces (PVS) are fluid filled channels between the vascular lumen, where blood flows, and the interstitial space where neurons reside (Wardlaw et al., 2020). During certain physiological states, such as deep sleep, the transport of fluids and solutes within the PVS are significantly increased (Xie et al., 2013), resulting in enhanced clearance of molecular waste from the nervous system (i.e., glymphatic clearance) (Iliff et al., 2012). Impairment of glymphatic clearance is associated with sleep deprivation and many disease states. It is hypothesised to exacerbate the accumulation of toxic metabolic byproducts in the brain, detrimental to neuronal health and cognitive function (Eide et al., 2021; Iliff, Lee, et al., 2013; Iliff, Wang, et al., 2013).

Enlarged PVS are implicated in multiple neurological diseases including cerebral small vessel disease (CSVD), ischaemic stroke, Alzheimer's disease, multiple sclerosis, and traumatic brain injury (Duering et al., 2023; Francis et al., 2019; Wardlaw et al., 2013). Enlarged PVS can be visualised on anatomical MRIs, thus allowing clinicians to assess changes in PVS associated with disease states and potentially disease progression and recovery.

Manual segmentation of PVS remains the gold standard for voxel-wise delineation, but it is time-consuming and labour-intensive, making it impractical for large-scale studies. Advancements in biomedical image processing have provided various approaches for the automated segmentation of PVS from anatomical MRI (Pham et al., 2022; Waymont et al., 2024). One of the most commonly used neural network architectures for biomedical image segmentation is the U-Net, which involves a contracting encoder pathway for feature extraction followed by an expanding decoder pathway which generates the final labelled mask (Ronneberger et al., 2015). Currently, the no-new-UNet (nnU-Net) is considered the benchmark for biomedical image segmentation (Isensee et al., 2021). The nnU-Net framework relies on the original U-Net architecture aided by rule-based selection of architecture hyperparameters, data handling, and preprocessing procedures (Isensee et al., 2021). The latest version called the nnU-Net Residual Encoder (ResEnc) incorporates residual connections within convolutional blocks during the encoder pathway consistently outperformed previous versions of the nnU-Net (Isensee et al., 2022).

To improve PVS visibility, image preprocessing such as image denoising techniques and contrast enhancement techniques are often used. To date, no attempts have been made to combine image denoising and contrast enhancement techniques with convolutional neural network-based PVS segmentation. In this study, we used the nnU-Net ResEnc, the latest version of the no-new U-Net, for automated PVS segmentation in T1-weighted (T1w) MRI. We applied a sparse annotation strategy to efficiently acquire PVS segmentation and investigate image preprocessing techniques (denoising with non-local means filtering [NLMF] and contrast adjustment with adaptive histogram equalisation [AHE]), iterative label cleaning, and a semi-supervised learning paradigm with pseudo-labels to improve segmentation performance. Importantly, we assessed the robustness of white matter (WM) and basal ganglia (BG) PVS segmentation when the model was applied to data obtained at different MRI field strengths. Lastly, we developed a model to include a fluid attenuated inversion recovery (FLAIR) sequence, enabling concurrent segmentation of white matter hyperintensities with PVS, and developed models dedicated to PVS segmentation in the less studied midbrain and hippocampal structures.

## 2. Methods

### 2.1. Primary datasets

#### *2.1.1. Participants*

Our primary training dataset comprised MRI data acquired from healthy controls (n=30) recruited as part of three separate studies. Each study involved different participant inclusion and exclusion criteria and imaging protocols. Ten participants were included from each of the three datasets. We refer to these three main datasets as Dataset A, Dataset B, and Dataset C. Dataset A was curated with ethics approval from the Human Research Ethics Committee (HREC) of the University of Sydney (Crawford et al., 2021). Dataset B was curated with ethics approval from the HRECs of the University of Western Sydney and the University of New South Wales (Fatouleh et al., 2015). Dataset C, also known as the Cognition and Neocortical Volume After Stroke (CANVAS) study, was approved by the respective ethics committees from three Stroke units at the Austin Hospital, Box Hill Hospital, and the Royal Melbourne Hospital (Brodtmann et al., 2014). In accordance with the Declaration of Helsinki, all participants provided informed written consent.

Participant demographics differed across the three MRI datasets. Participants in Dataset A were the youngest (5 women, mean±SD age=28.3±11.5 years) compared to Dataset B (5 women, mean age=55.5±10.4 years) and Dataset C (3 women, mean age=66.1±8 years).

### *2.1.2. MRI acquisition*

T1-weighted (T1w) MPRAGE sequence images were curated from the three datasets. Dataset A used images acquired on a 7T Siemens, Magnetom model MRI scanner with a single-channel transmit and 32-channel receive head coil (Nova Medical) (TR = 5000 ms, TE = 3.1 ms, FA = 4°, FOV = 240 mm × 167.25 mm, voxel size = 0.75 mm isotropic). Dataset B was acquired on a 3T Philips, Achieva model MRI scanner with a 32-channel SENSE head coil (TR = 5600 ms, TE = 2.5 ms, FA = 8°, FOV = 250.56 mm × 174 mm, voxel size = 0.87 mm isotropic). Dataset C was acquired on 3T Siemens, Trio Tim model MRI scanners with a 12-channel head coil (Siemens, Erlangen, Germany). This dataset included acquisition of a T1-weighted sequence (TR = 1900 ms, TE = 2.55 ms, TI = 900 ms, FA = 9°, FOV = 256 mm × 256 mm, voxel size = 1.0 mm isotropic), a T2-weighted (T2w) sequence (TR = 3390 ms, TE = 390 ms, FA = 120°, FOV = 256 mm × 256 mm, voxel size = 1.0 mm isotropic), and a FLAIR sequence (TR = 6000 ms, TE = 380 ms, TI = 2100 ms, FA = 120°, FOV = 512 mm × 512 mm, voxel size = 1.0 mm isotropic). Supplementary Table 1 provides the MRI parameters of these three datasets. FLAIR images in Dataset C were used to differentiate white matter hyperintensities (WMH) from perivascular spaces. All available images from the healthy participants were visually inspected. From each dataset, ten subjects with high PVS burden were selected for manual segmentation. Poor quality images due to motion artefacts, ghosting, Gibbs ringing, or phase-encoding artefacts were avoided.

## 2.2. Secondary datasets

A large secondary dataset was derived from 17 open-access repositories consisting of 12,605 images and the CANVAS study consisting of 135 images. MRI scans were acquired from healthy controls (n=7,269, $n_{women}$=5,886, mean age=49.3±24.9 years) and patients diagnosed with neurological conditions such as ischaemic stroke (n=756), mild cognitive impairment (n=881), Alzheimer’s disease (n=392), older individuals at risk of Alzheimer’s disease (n=1,750), Parkinson’s disease (n=668), autism spectrum disorder (n=385), attention deficit hyperactivity disorder (n=157), traumatic brain injury (n=310), and frontotemporal dementia (n=172). The open-access datasets included the Anti-Amyloid Treatment in Asymptomatic Alzheimer's Disease (A4) study, Autism Brain Imaging Data Exchange (ABIDE), Aging Brain: Vasculature, Ischemia, and Behavior Study (ABVIB), ADHD-200, Alzheimer's Disease Neuroimaging Initiative (ADNI-2 and 3), Alzheimer's Disease Neuroimaging Initiative Department of Defense (ADNIDOD), Australian Imaging Biomarkers and Lifestyle (AIBL) study, Amsterdam Open MRI Collection (AOMIC ID1000), Anatomical Tracings of Lesions

After Stroke (ATLAS-2), Consortium for Reliability and Reproducibility (CoRR), Human Connectome Project (HCP1200), I See Your Brain (ISYB), frontotemporal lobar degeneration neuroimaging initiative (FTLDNI/NIFD), Nathan-Kline Institute Rockland Sample (NKI-RS), Open Access Series of Imaging Studies (OASIS-3), and Parkinson's Progression Markers Initiative (PPMI) (Aleksovski et al., 2018; Bellec et al., 2017; Di Martino et al., 2014; Ellis et al., 2009; Gao et al., 2022; Glasser et al., 2013; Jack et al., 2008; LaMontagne et al., 2019; Liew et al., 2022; Nooner et al., 2012; Petersen et al., 2010; Snoek, van der Miesen, Beemsterboer, et al., 2021; Snoek, van der Miesen, van der Leij, et al., 2021; Van Essen et al., 2013; Veitch et al., 2023; Weiner et al., 2017; Zuo et al., 2014).

Participant data were included if a baseline MRI scan with field strength≥3T and slice thickness ≤1.2 mm was available. The secondary datasets were used to: i) improve performance with a semi-supervised learning strategy (i.e., pseudo-labels) (Section 2.4.5), ii) supplement the training dataset in the midbrain (MB) or hippocampus (HP) PVS segmentation tasks and improve the likelihood of successful model convergence after training (Section 2.5), and iii) supplement the training dataset in the T1w+FLAIR and T2w nnU-Net models (Section 2.6-2.7). They also provided data from participants with a range of ages and disease states, boosting potential real-world applicability.

Pseudo-labelling is a recent technique developed to boost model performance without the need for additional manual annotation of data, offering a low-cost solution to optimising model performance (Ferreira et al., 2023). One baseline MRI scan per participant was selected for pseudo-labelling. All T1w images (n=12,740) from the secondary datasets were automatically segmented for WM and BG-PVS (Section 2.4.4), resulting in 12,740 pseudo-labels. These were included in the training dataset for an improved nnU-Net model (Section 2.4.5). Supplementary Figure 1 and Tables 2-5 provide demographic details, descriptive statistics of PVS quantities detected, and MRI parameters for the secondary datasets used for pseudo-labelling. Images from the ADNI3, AIBL, and OASIS-3 datasets were used to train the nnU-Net models for MB and HP-PVS segmentation (Section 2.5). Demographic details and MRI parameters for images used to develop the MB-PVS model are described in Supplementary Table 6-7. Demographic details and MRI parameters for images used to develop the HP-PVS model are described in Supplementary Table 8-9.

Images from the AIBL, CANVAS, NIFD and PPMI datasets were used for the T1w+FLAIR nnU-Net model (Section 2.6), while images from CANVAS, HCP1200, and NIFD were used for the T2w nnU-Net model (Section 2.7). Demographic details and MRI parameters of images used to train the T1w+FLAIR model are described in Supplementary Tables 10-13. Demographic details and MRI parameters of T1w images used to train the T2w model are described in Supplementary Tables 14-16. Figure 1 illustrates the overall workflow of the study.

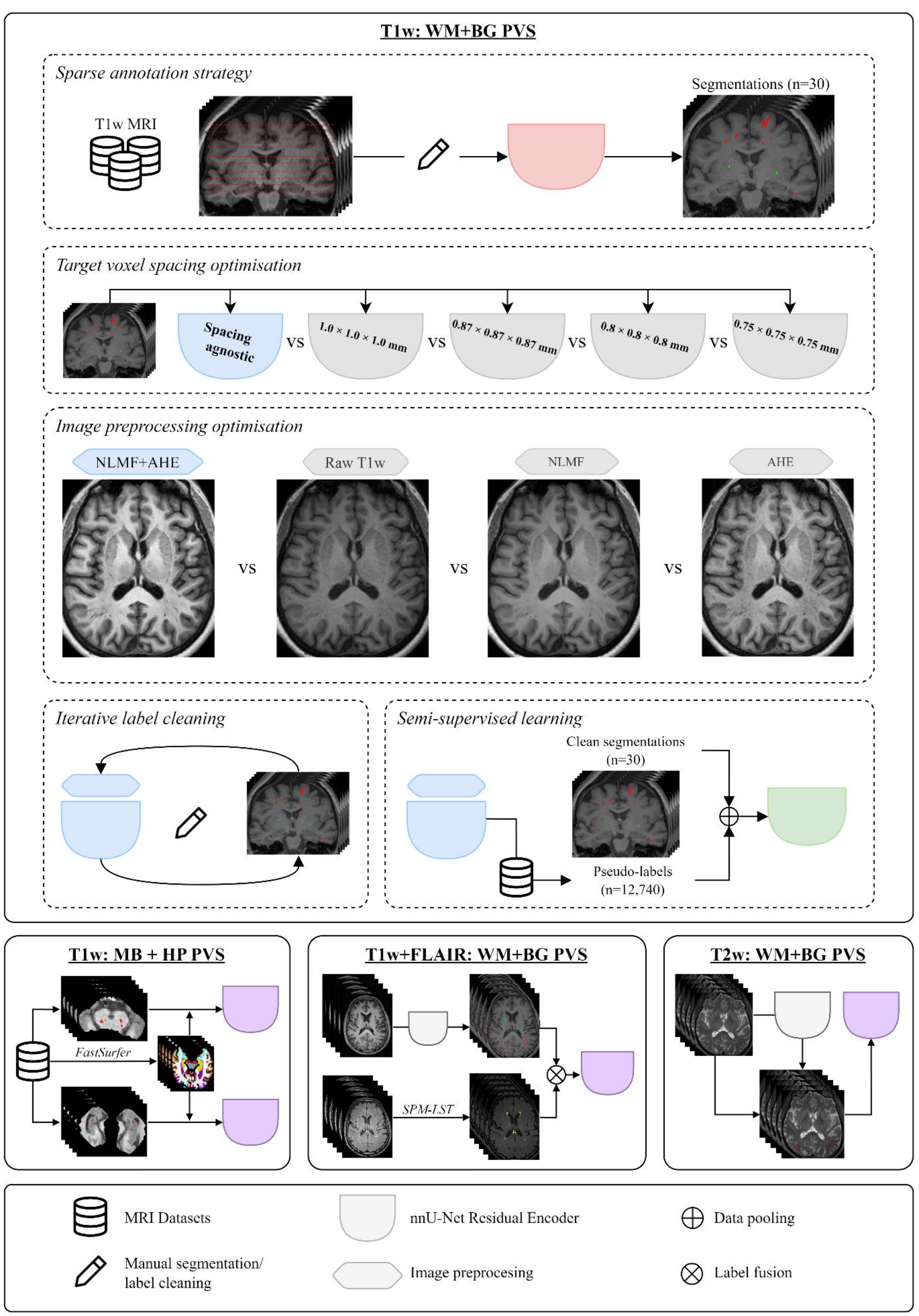

**Figure 1.** Overall workflow of the study. Panel 1: development of the nnU-Net PVS segmentation in the white matter (WM) and basal ganglia (BG) using T1-weighted (T1w) MRI. Sub-panels show the acquisition of initial labelled data with a sparse annotation strategy, followed by model optimisation by comparing image handling parameters, preprocessing techniques, iterative label cleaning, and semi-supervised learning with pseudo-labels. Panel 2: development of nnU-Net models for PVS segmentation in the midbrain (MB) and hippocampi (HP). Panel 3: development of nnU-Net models using T1w and FLAIR images to segment WM and BG-PVS alongside white matter hyperintensities (WMHs). Panel 4: development of models using T2w images for segmentation of WM and BG-PVS. NLMF: non-local means filtering. AHE: adaptive histogram equalisation.

Data used in the preparation of this article were obtained in part from the Alzheimer's Disease Neuroimaging Initiative (ADNI) database (adni.loni.usc.edu) and the Aging Brain: Vasculature, Ischemia, and Behavior Study database (ABVIB). The ADNI was launched in 2003 as a public-private partnership, led by Principal Investigator Michael W. Weiner, MD. The ABVIB study was launched in 1994 as a NIA-funded program project led by Principal Investigator Helena C. Chui, MD. Data from AIBL was collected by the AIBL study group. AIBL study methodology has been reported previously (Ellis et al., 2009). NIFD/FTLDNI was funded through the National Institute of Aging and started in 2010. The Principal Investigator of NIFD was Dr. Howard Rosen, MD at the University of California, San Francisco. The data are the result of collaborative efforts at three sites in North America. For up-to-date information on participation and protocol, please visit http://memory.ucsf.edu/research/studies/nifd.

## 2.3. Model architecture

All nnU-Net models were based on the 3D full-resolution configuration with the Residual Encoder or 'ResEncM' variant. For simplicity, we refer to the nnU-Net Residual Encoder as the nnU-Net, not to be confused with the 'plain-vanilla' version which does not implement residual connections in the encoder pathway (Isensee et al., 2021, 2024). The ResEnc variant is the most recent update to the nnU-Net pipeline and offers substantially improved segmentation performance compared to the 'plain-vanilla' model.

The nnU-Net framework allows for the customisation of image handling and preprocessing parameters. By default, the target voxel spacing of the nnU-Net is the median spacing of the training images. We tested customised voxel spacings and a voxel spacing agnostic configuration (Section 2.4.2). The voxel spacing agnostic model assumes all images were of the same resolution, thereby avoids resampling of images during preprocessing and inference. The default image preprocessing involves voxel intensity clipping between the 0.5 and 99.5 percentiles followed by z-score normalisation. All models used the default preprocessing except for the variations that used adaptive histogram equalisation (Section 2.4.3) and did not apply z-score normalisation. Data augmentation was conducted with the 'batchgenerators' package, which applies various spatial augmentations including mirroring, elastic deformation, image rotations, and rescaling, as well as noise augmentation with the addition of gaussian noise (Isensee et al., 2020).

## 2.4. T1w nnU-Net: WM and BG-PVS

### *2.4.1. Sparse annotation strategy*

Ten T1w MRI scans from healthy controls were selected from each of the main datasets (A/B/C). For each image, PVS were manually labelled in only ten axial slices by a single rater (WP) (Figure 2). A nnU-Net model was then employed to complete the segmentations using a sparse annotation strategy. Axial slices that were not manually labelled were assigned an 'ignore' label and models were trained with the partial loss function (Gotkowski et al., 2024). These PVS segmentations (n=30) were reviewed and manually corrected as required. Subsequently, we assigned different PVS voxel-labels for in various brain regions including white matter (WM), assigned a '1' label, and basal ganglia (BG), assigned a '2' label. All manual segmentations were conducted with ITK-Snap (Yushkevich et al., 2006). The segmentations (n=30) were then used to train and optimise a single nnU-Net pipeline.

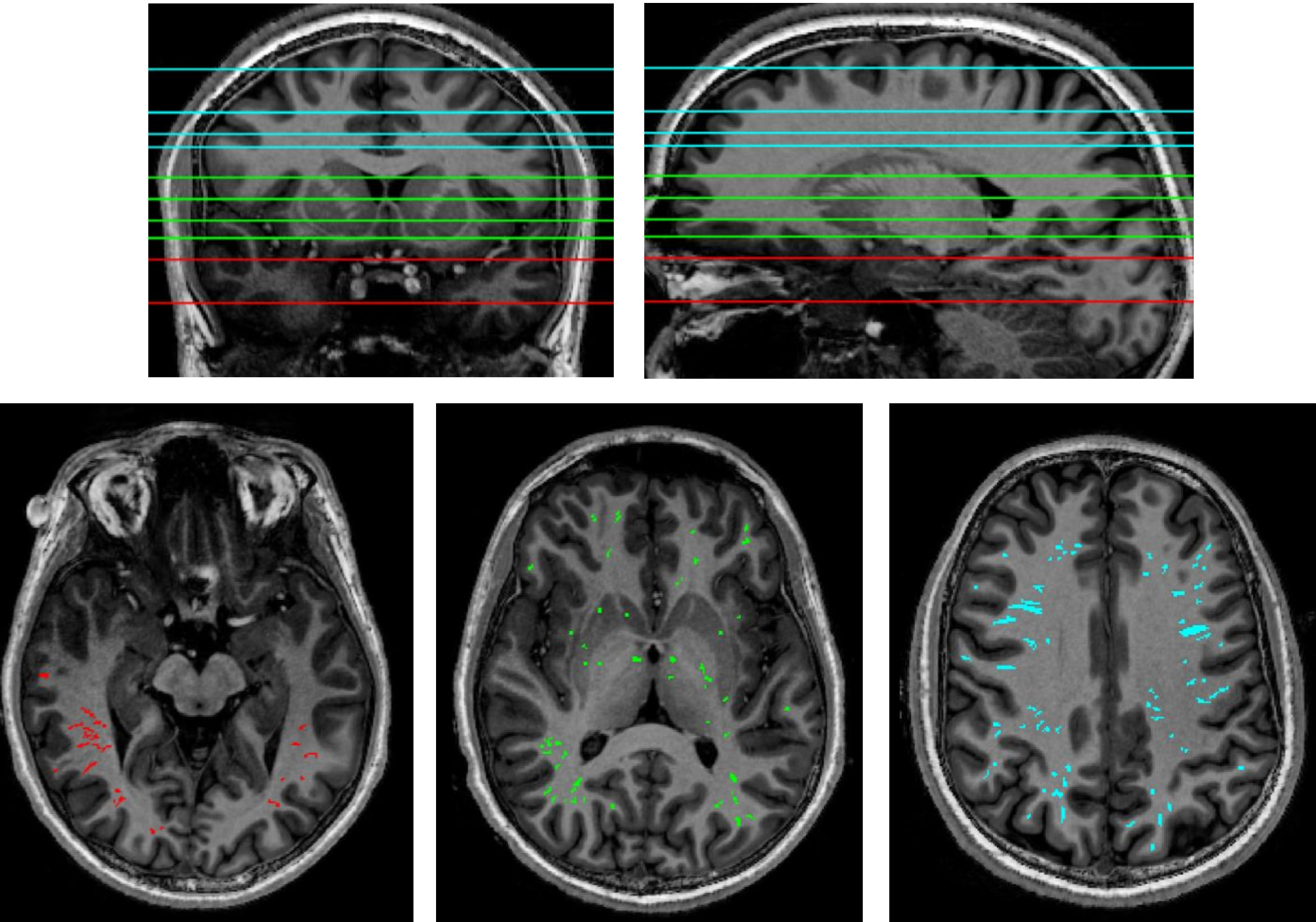

**Figure 2.** T1-weighted MRI demonstrating the sparse segmentation strategy. 10 Axial slices from each image were selected for manual segmentation (Top-Left: coronal slice; Top-Right: sagittal slice). Red bands indicate the two axial slices at the brainstem (BS) level. Green bands indicate the four axial slices selected at the basal ganglia (BG) level. Cyan bands indicate the four axial slices selected at the centrum semiovale (CS) level. Bottom: axial slices with MRI-PVS labelled.

### 2.4.2. *Target spacing optimisation*

We trained and tested nnU-Net models on all 30 labelled images to determine a robust image handling pipeline. First, the effects of various target voxel spacings were benchmarked against each other. Four different target spacings were compared. This included 0.75-, 0.80-, 0.87- and 1.00-mm isotropic voxel spacing. Additionally, a voxel spacing agnostic model, which ignored voxel spacing information and did not resample images to a common spacing, was benchmarked against the previous configurations. The most performant target spacing was used in all subsequent models. The configurations were validated in a stratified 5-fold cross-validation (5FCV).

### 2.4.3. *Image preprocessing optimisation*

Next, we evaluated the effects of two image preprocessing techniques on segmentation performance. Our preprocessing pipeline involved several steps: background removal with Otsu thresholding, voxel intensity rescaling between 0 and 1, followed by denoising with NLMF, and contrast adjustment with AHE (Coupe et al., 2008; Coupé et al., 2012). Image preprocessing was implemented in Python (v3.9). NLMF was implemented via the 'dipy' package (Garyfallidis et al., 2014). Otsu thresholding and AHE were implemented via the 'scikit-image' package (Walt et al., 2014). The voxel-spacing agnostic nnU-Net was trained on T1w images processed with NLMF only, AHE only, and combined NLMF+AHE (Figure 3). The effect of image preprocessing methods was assessed with 5FCVs.

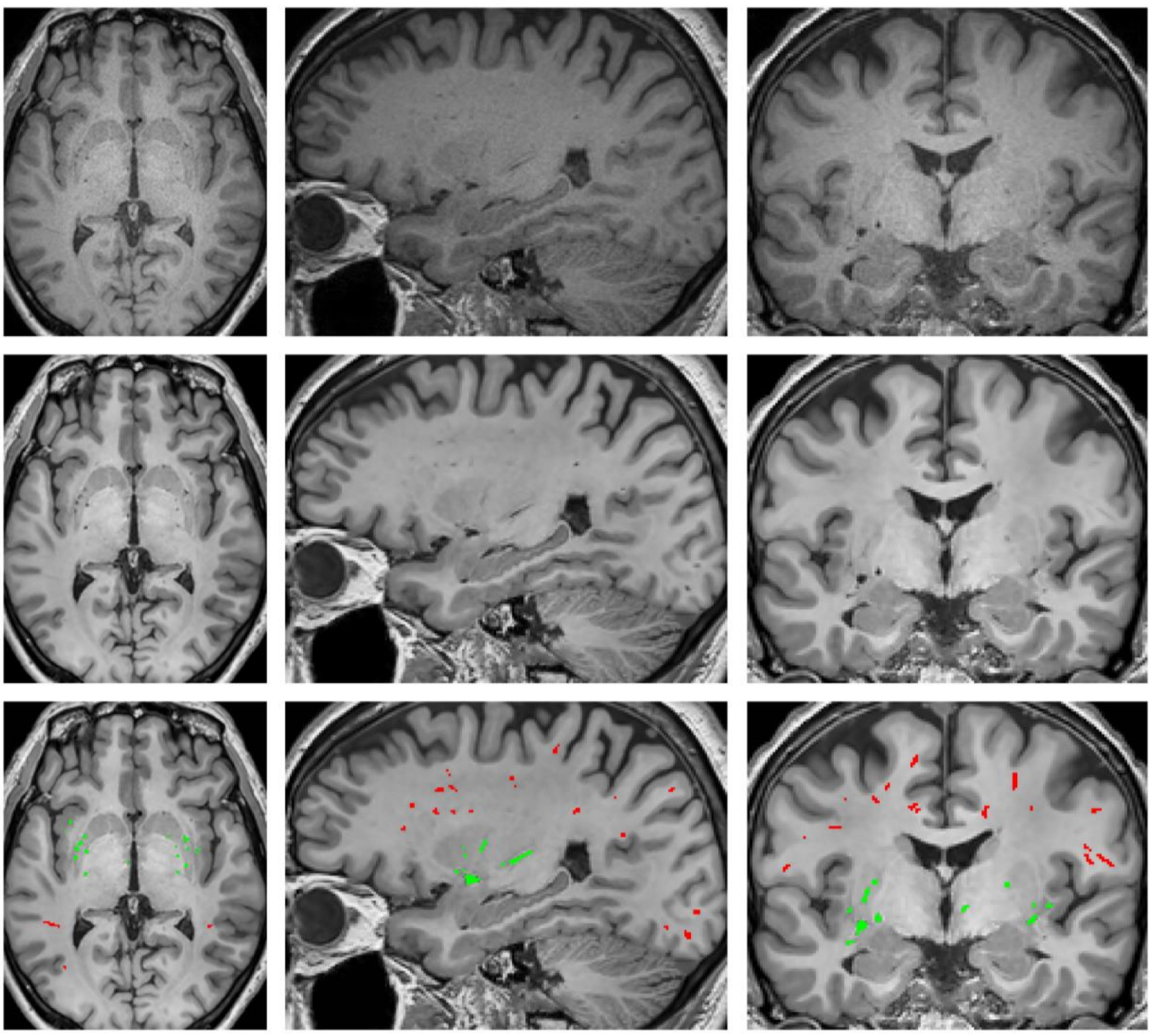

**Figure 3.** The effect of image enhancement on a T1w scan from Dataset B (3T, Philips Achieva, 0.87 mm isotropic voxel size). Top: raw T1w image. Middle: T1w image after non-local means filtering (NLMF) and adaptive histogram equalisation (AHE). Bottom: manual PVS segmentation overlay with WM-PVS (red) and BG-PVS green.

### *2.4.4. Iterative label cleaning*

The combination of both NLMF and AHE greatly improved PVS visibility. Therefore, we used the enhanced images for manual cleaning of the training data. Manual segmentation was conducted by a single rater (WP) then reviewed by an experienced radiologist (AJ). This comprised two rounds of model training and manually quality checking of PVS segmentations. The voxel spacing agnostic model incorporating both NLMF and AHE (enhanced nnU-Net) was retrained and validated in a 5FCV to determine the effect of iterative label cleaning on model performance.

### 2.4.5. Semi-supervised learning

We subsequently used pseudo-labels (model predictions generated on unseen and out-of-sample images) to increase the size of the training dataset and improve the nnU-Net model. The enhanced nnU-Net was applied to T1w images in the secondary datasets. Subsequently, the voxel spacing agnostic nnU-Net with NLMF and AHE was retrained in a 5FCV with pseudo-labels (n=12,740) included as training data.

Lastly, we assessed the final T1w model against the original ground truth segmentations derived from raw T1w MRI scans (Section 2.4.1). The pipeline with image preprocessing and enhancement was applied to the T1w images resulting in 30 predictions. Predictions were compared against the initial ground truth segmentations (Section 2.4.1). Standard segmentation performance metrics were computed. Correlation and concordance statistics were computed to evaluate the agreement between predicted and actual PVS voxel and cluster counts across the whole brain, white matter and BG.

## 2.5. T1w nnU-Net: MB and HP-PVS

We optimised two nnU-Net models for the segmentation of PVS in the midbrain (MB) and hippocampal (HP) regions using T1w MRI. The MB-PVS nnU-Net was trained using images from 60 participants (28 women, mean age=59.7±17.3 years): 30 from the primary datasets, 10 from ADNI3, 10 from AIBL, and 10 from OASIS-3. The HP-PVS nnU-Net was trained using 50 images (20 women, mean age=68.2±10.1 years): 10 from each of the datasets B, C, ADNI3, AIBL, and OASIS-3. Dataset A was not included for HP-PVS segmentation as they did not contain observable HP-PVS. MB and HP-PVS were identified by a single rater (WP) and reviewed by an experienced radiologist (AJ) (Figures 4-5) according to established guidelines (Adams et al., 2013, 2015; Potter et al., 2015).

We tested three different variations of the nnU-Net for PVS segmentation. The first variation was trained using raw T1w MRI scans as the input. The second variation included a brain parcellation mask as the secondary input channel. Parcellation masks consisted of a unique integer label assigned to each brain region according to the Desikan-Killiany-Tourville atlas and derived by FastSurfer (v2) (Henschel et al., 2020). The third variation involved a single input channel with a post-processed T1w MRI scan where all brain regions external to the midbrain or hippocampus were removed from the image (Figures 4-5). We report results only from the third variation as it is the only one that converged during the 1000-epoch model training in each of the training folds.

In this last variation, anatomical structures outside the regions of interest (ROI), midbrain and hippocampi, were removed in the T1w images. FastSurfer (v2) brain parcellation masks, registered to the Desikan-Killiany-Tourville atlas, were combined with the raw T1w MRI scans (Desikan et al., 2006; Henschel et al., 2020). For the midbrain-PVS model, the ROI designated as the 'ventral-diencephalon' was retained in the T1w images. For the hippocampus-PVS model, the ROIs designated as 'left-hippocampus' and 'right-hippocampus', were retained. Binary masks of the ROIs were dilated using the 'binary_dilation' function from the 'scipy.morphology' (v.1.6.2) Python package to ensure all masks covered the entirety of the ROIs.

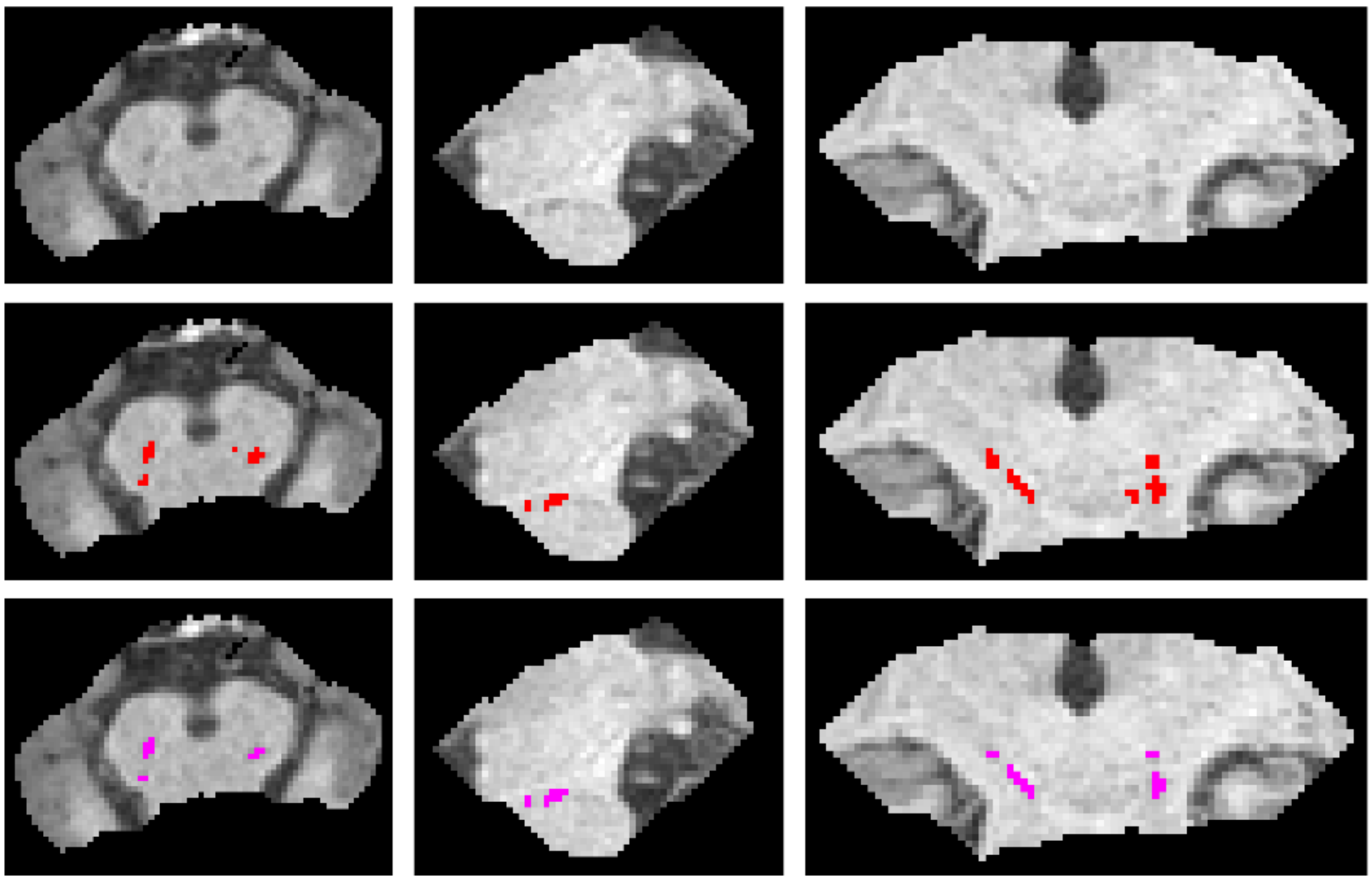

**Figure 4.** Axial slices from a T1w scan showing PVS labelled in the midbrain. Top row: raw images. Middle row: manually labelled PVS in red. Bottom row: automatically labelled PVS in magenta. Left column: axial plane. Middle column: sagittal plane. Right column: coronal plane.

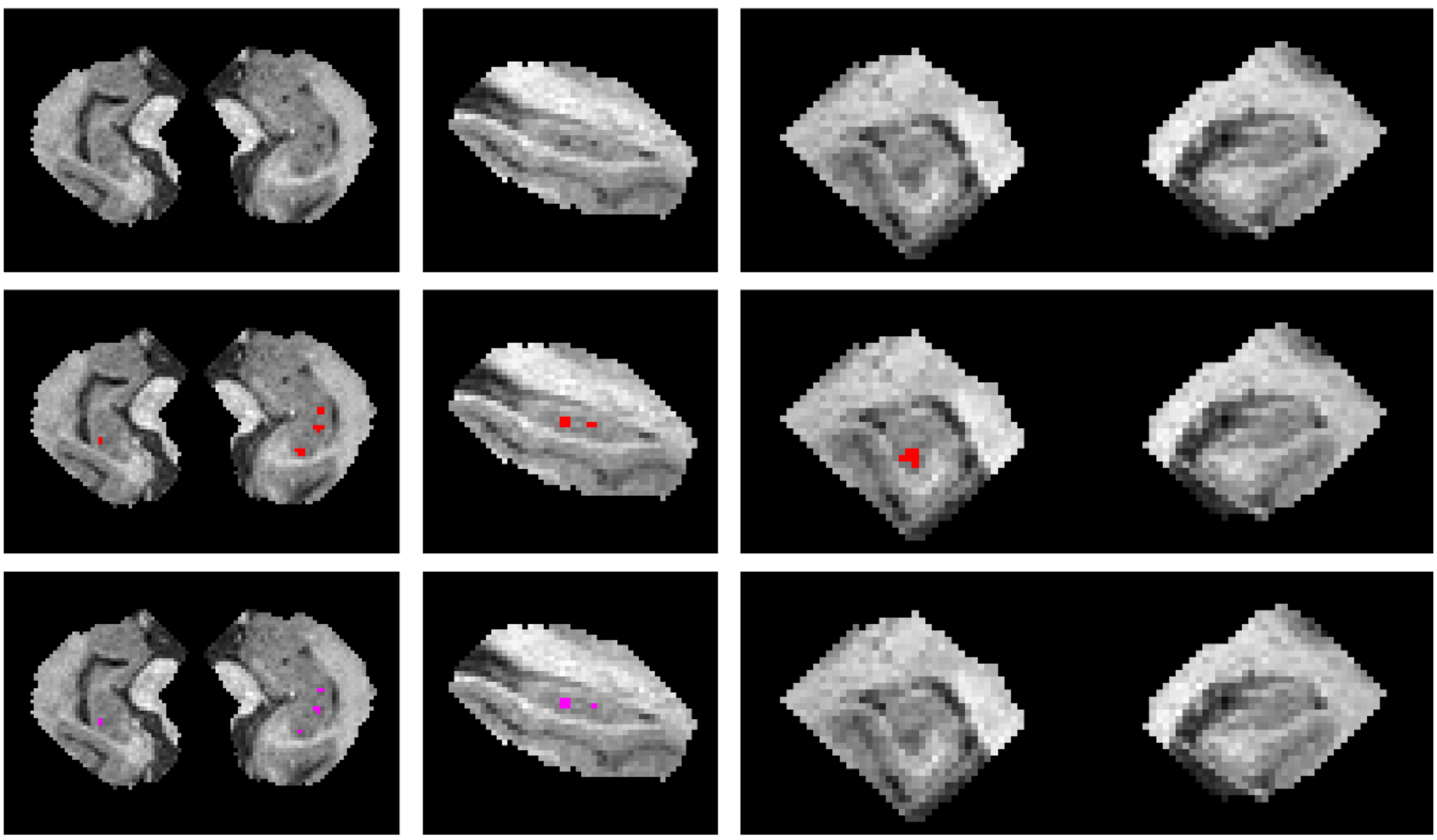

**Figure 5.** Axial slices from a T1w scan showing PVS labelled in the hippocampi. Top row: raw images. Middle row: manually labelled PVS in red. Bottom row: automatically labelled PVS in magenta. Left column: axial plane. Middle column: sagittal plane. Right column: coronal plane.

2.6. T1w+FLAIR nnU-Net

We trained a dual-channel voxel-spacing agnostic nnU-Net with 3D T1w MRI scans as the primary input and co-registered 3D FLAIR scans as the secondary input. The T1w+FLAIR model was trained on a subset of the secondary datasets. This included 958 images from the AIBL (n=310), CANVAS (n=135), NIFD (n=301), and PPMI (n=213) datasets. One image randomly selected from each dataset was withheld for model testing. The T1w pseudo-labels (Section 2.4.5) for these held-out images were combined with white matter hyperintensity (WMH) masks generated by the Lesion Segmentation Tool Lesion Prediction Algorithm (Schmidt et al., 2019). This resulted in segmentations with unique voxel-labels for the WM-PVS, BG-PVS, and WMHs (Figure 6). WM-PVS were assigned the ‘1’ label, BG-PVS were assigned the ‘2’ label, and WMHs were assigned the ‘3’ label. Subsequently, the four corresponding segmentations (Section 2.4.5) were manually corrected and used to train an initial T1w+FLAIR model. The model was applied to the remaining image pairs to generate pseudo-labels (n=995). A final T1w+FLAIR model was trained on the pseudo-labels (n=995) and tested on the held-out image pairs (n=4).

FLAIR images were registered to their respective T1w images using the 'ANTSRegistrationSyn' algorithm with rigid and affine registration, and nearest neighbour interpolation (Avants et al., 2008, 2011). Lesion Prediction Algorithm in SPM12-MATLAB (v.r2018a) was used to generate the WMH lesion masks. The masks were binarised based on a 0.5 threshold, using 'fslmaths' function from the FSL toolkit (v6.0.4) (Jenkinson et al., 2012). The WMH masks were used to identify and exclude false PVS labels.

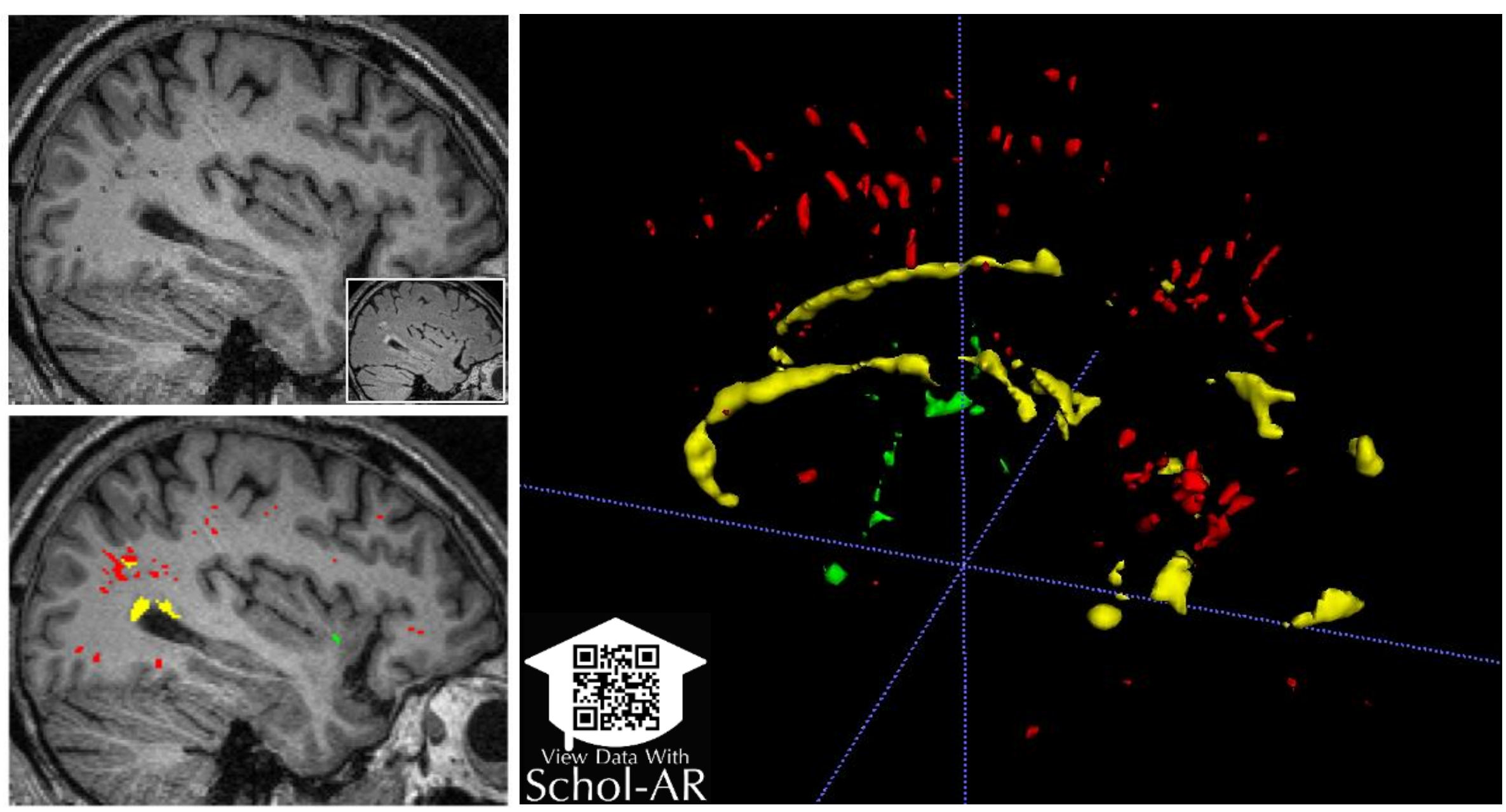

**Figure 6.** PVS segmentation from the NIFD dataset used to test the T1w+FLAIR model. Top-left: Raw T1w scan with corresponding FLAIR scan (inset). Bottom-left: PVS segmentation generated by the T1w+FLAIR nnU-Net overlaid onto T1w scan with WM-PVS (red), BG-PVS (green), and white matter hyperintensities (yellow). Right: 3D render of the PVS segmentation is augmented with Schol-AR and viewable via the QR code (Ard et al., 2022).

2.7. T2w nnU-Net

We used T2w images (n=1,448) from the CANVAS (n=132), HCP1200 (n=1,104), and NIFD (n=212) datasets to train and optimise a nnU-Net for WM and BG-PVS segmentation. The pseudo-labels generated with T1w MRI were reused for the T2w MRI nnU-Net model. As with the FLAIR scans, T2w MRI scans were registered to the T1w scans using the 'ANTSRegistrationSyn' algorithm with rigid and affine registration and nearest neighbour interpolation (Avants et al., 2008, 2011). One image randomly selected from each dataset was withheld for model testing. The three corresponding pseudo-labels (Section 2.4.5) were manually corrected and used to train an initial T2w model. The model was applied to the remaining T2w scans to generate pseudo-labels (n=1,445). A final T2w model was trained on the pseudo-labels (n=1,445) and tested on the held-out T2w images (n=3).

### 2.8. Model training

Models were trained with a combination of the dice and cross-entropy loss function:

$$\mathcal{L}_{total} = \mathcal{L}_{dice} + \mathcal{L}_{CE} \quad (1)$$

To facilitate multi-class labelling our dice loss formulation was based on a multi-class adaptation proposed previously (Isensee et al., 2022):

$$\mathcal{L}_{dc} = -\frac{2}{|K|}\sum_{k\in K}\frac{\sum_{i\in I} u_i^k v_i^k}{\sum_{i\in I} u_i^k + \sum_{i\in I} v_i^k} \quad (2)$$

Where $u$ refers to the softmax output of the network and $v$ refers to the one-hot encoding of ground truth segmentation labels. $u$ and $v$ have the shape $I \times K$. $i \in I$ represents the number of voxels in the training patch and batch. $k \in K$ represents the number of label classes.

Models were trained with the Adam optimiser via an initial learning rate = $3 \times 10^{-4}$ and batch size of two (Kingma & Ba, 2017). Each epoch comprised an iteration over 250 training batches. An exponential moving average of the training loss are computed during training. If the exponential moving average of the training loss did not improve by at least $5 \times 10^{-3}$ within the last 30 epochs, then the learning rate was decreased by a factor of 5. Each training fold consisted of 1000 training epochs.

All models were developed with the nnU-Net v2.4.2 in Python (v3.9) and PyTorch (v2.3), supported by CUDA (v11.8). Models were trained on a Nvidia V100 or A40 GPUs with 250GB RAM, Intel Xeon Gold 5320, 13 cores, using the M3 High-Performance Computing cluster (Goscinski et al., 2014).

### 2.9. Model evaluation

All T1w models were evaluated in a stratified 5FCV procedure. The training dataset was partitioned into five equally sized splits. In each fold, the model was trained on four splits and validated on a held-out split. Images were stratified such that each split included an equal number of images per dataset. We ensured an equal mix of high PVS and low PVS images in all training folds and in the validation set. To ensure comparability, T1w models targeting WM and BG-PVS were validated using data splits that were preserved across all cross-validation procedures (Section 2.4). For the model trained with pseudo-labels, all pseudo-labels were included alongside the training data in each training fold of the 5FCV.

The T1w models targeting midbrain-PVS and hippocampus-PVS were also evaluated in a stratified 5FCV. Since the MB-PVS and HP-PVS segmentations relied on different images, the 5FCV was performed independent of each other. Each data split included an equal number of images per dataset. The model targeting MB-PVS was evaluated with 60 images. The model targeting HP-PVS was evaluated with 50 images.

The T1w+FLAIR and T2w models were evaluated via a train/test procedure. There were insufficient manual segmentations to conduct a 5FCV. For both models, a single image per dataset was held out for testing with model training performed on the remaining images. The T1w+FLAIR model was trained on images (n=955) from four datasets and tested on four images. The T2w model was trained on images (n=1,445) from three datasets and tested on three images.

*2.9.1. Evaluation metrics*

For performance evaluation, the nnU-Net package computes the DSC, the number of true positives, false positives, and false negatives. Additionally, we computed the sensitivity (SEN) and positive predictive value (PPV). These metrics are determined based on voxel-wise comparisons between the ground truth reference segmentations and model predictions. The DSC is a measure of overall segmentation performance. Sensitivity evaluates the proportion of PVS voxels correctly detected. Positive predictive value evaluates the proportion of predicted voxels that were correctly labelled as PVS.

$$DSC = \frac{2TP}{(TP+FP)+(TP+FN)},\ SEN = \frac{TP}{(TP+FN)},\ PPV = \frac{TP}{(TP+FP)} \qquad (3)$$

To assess model performance with respect to cluster-wise measurements, we complemented the voxel-wise metrics with Lin's concordance correlation coefficient (CCC), Bland-Altman plots, and Spearman's correlation tests. These tests were used to compare the number of predicted PVS clusters to the actual number of clusters present in the manual labels. Lin's CCC evaluates the agreement between predicted and actual values. Spearman's correlation test evaluates the trend between predicted and actual values.

All quantitative metrics were evaluated for each label in their respective models. This included the WM-PVS and BG-PVS labels in the T1w and T2w models. In the T1w+FLAIR model, the segmentation performance of the WMH labels was also evaluated. Similarly, all performance metrics were assessed for the MB-PVS and HP-PVS models. Model predictions were visually compared to their respective manual segmentations at appropriate stages throughout the model development process.

### *2.9.2. Statistical analysis*

Descriptive statistics of PVS metrics including voxel and cluster counts are presented in Tables 1-5. Boxplots were used to detect outliers in the training datasets. The normality of PVS metrics and model performance metrics were tested via Shapiro-Wilk tests and inspection of Q-Q plots. Spearman's non-parametric correlation tests were conducted. Lin's CCC with 95% confidence interval (CI) were estimated with the 'epiR' (v2.0.74) package. The statistical significance level was set at 0.05. Statistical analyses and figure generation was performed using R (v4.3.3) and R Studio (v2023.12.1-402).

# 3. Results

## 3.1. T1w nnU-Net: WM+BG-PVS

### *3.1.1. Sparse annotation strategy*

Initial sparse segmentations produced by model predictions consisted of mean±SD PVS voxels and cluster counts of 6812.5±3922.75 and 286±139.75, respectively. Eight of the ten corresponding FLAIR scans in dataset C showed WMHs confluent with PVS. Some voxels mistakenly labelled as PVS were part of adjoining cortical grey matter which became evident upon reviewing the sagittal and coronal planes.

### 3.1.2. *Target spacing optimisation*

There was a noticeable trend between target spacing and model performance, evaluated by 5FCV DSC scores. In general, performance was higher for models with higher resolution target spacing (smaller voxel size). Compared to alternative target spacings, the highest resolution model trained at 0.75 mm isotropic voxel spacing performed the best (mean±SD DSC=55±3%), followed by the 0.87 mm (DSC=49.2±1.9%), 0.80 mm (DSC=46.2±3.1%), and 1.0 mm (DSC=40.5±2.8%) isotropic voxel spacing (Table 1). Notably, the voxel-spacing agnostic model substantially outperformed all models that considered voxel-spacing in the image preprocessing pipeline (mean DSC=64.3±3.3%, WM-PVS DSC=70±1.2%, and BG-PVS=58.6±5.7%). Visual inspection of PVS labels generated by the voxel-spacing agnostic model further reinforced the robust segmentation performance in both WM and basal ganglia regions (Figure 7). Amongst the different models, PVS metrics from the voxel spacing agnostic model had the highest correlation and concordance with PVS voxel counts (ρ=0.94, Lin's CCC=0.79, 95%CI: 0.65–0.87) and cluster counts (ρ=0.87, CCC=0.86, 95%CI: 0.75–0.93) from the manual segmentations.

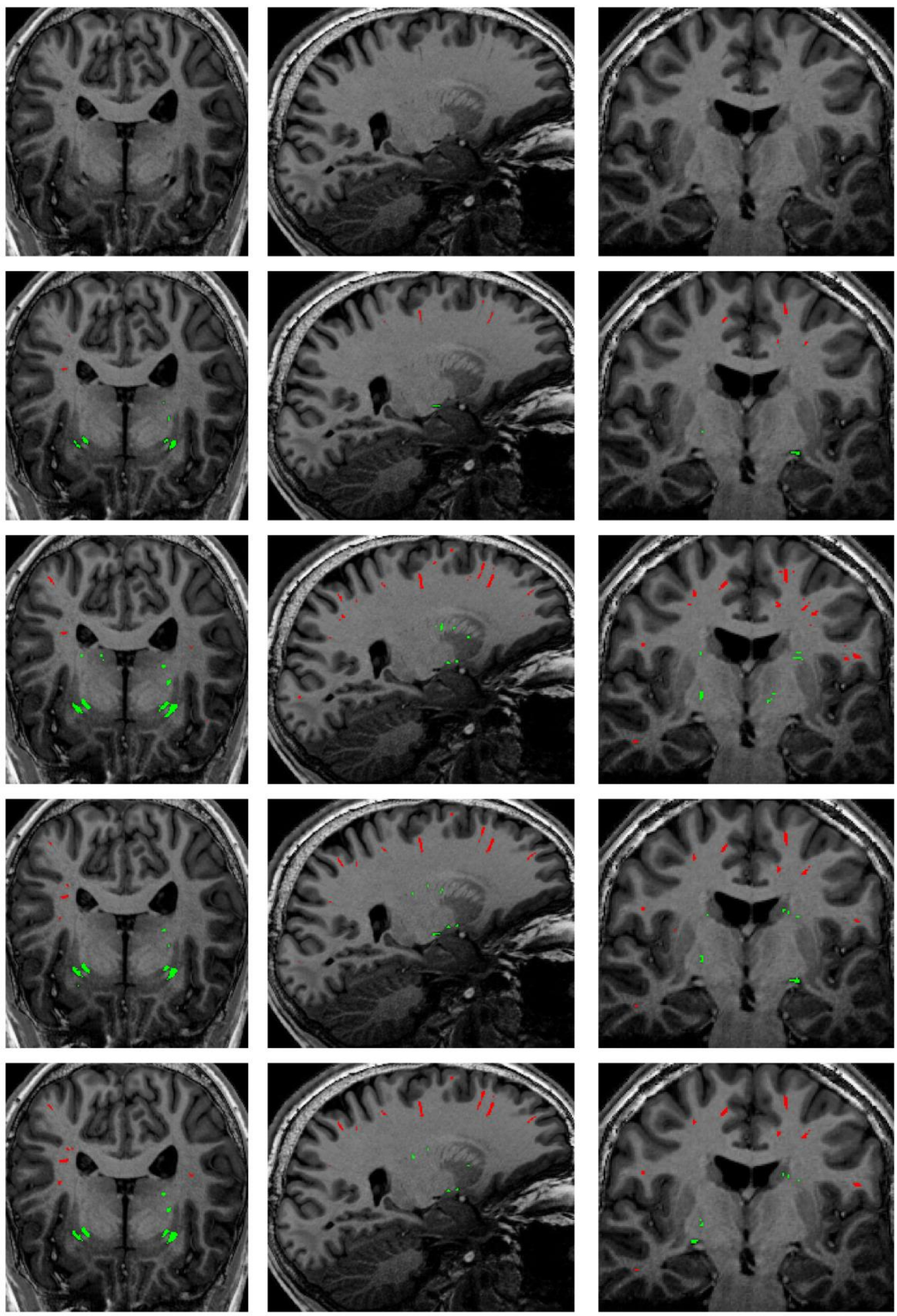

**Figure 7.** Comparisons of PVS segmentation from model prediction to manual from Dataset A (7T, Siemens Magnetom, 0.75 mm isotropic voxel spacing. Row 1: raw T1w image. Row 2: manual PVS segmentation overlay on T1w image. Row 3: prediction generated from the voxel-spacing agnostic nnU-Net. Row 4: prediction generated from the nnU-Net with 0.75 mm isotropic target spacing. Row 5: prediction generated form the nnU-Net with 1.0 mm isotropic target spacing. White matter PVS are labelled red. Basal ganglia PVS are labelled green.

#### 3.1.3. *Image preprocessing optimisation*

Next, we evaluated the effect of image processing on model performance. Trained with raw T1w MRI, the 5FCV DSC score for the spacing agnostic nnU-Net was 64.3±3.3%. Image preprocessing with NLMF alone resulted in a DSC=63.6±3.2% whereas AHE alone resulted in a DSC=63.4±3.1%. The model trained with T1w images processed with both NLMF and AHE yielded DSC=63±3.1% (Table 1). Notably, the voxel-spacing agnostic model trained with raw T1w images outperformed models trained with pre-processed images. Amongst the models using different preprocessing techniques, the model with both NLMF and AHE preprocessing had the highest concordance for PVS voxel counts (ρ=0.93, CCC=0.80, 95%CI: 0.68–0.88). Whilst the model with unprocessed images had the highest concordance for PVS cluster counts (ρ=0.87, CCC=0.86, 95%CI: 0.75–0.93).

#### 3.1.4. *Iterative label cleaning*

After iterative label cleaning, the number of PVS voxels and clusters labelled per image was reduced. The final PVS segmentations (n=30) consisted of median±IQR PVS voxels and cluster counts of 4296 ±3793.3 and 212±137.5, respectively. The median WM-PVS and BG-PVS voxels was 5669±3933.8 and 1150±376.5, respectively. The median number of PVS clusters in the WM and BG was 243±120.3 and 38.5±13.8, respectively. After each round of iterative label cleaning, the mean 5FCV DSC score increased to 80.8±0.9% and 85.7±1.2%. With respect to the WM-PVS and BG-PVS, the nnU-Net trained with final segmentations yielded 5FCV DSC scores of 89.5±1% and 82±2.5%. After iterative labelling was applied, the concordance between predicted and actual PVS voxel counts (ρ=0.97, CCC=0.98, 95%CI: 0.96–0.99) and PVS cluster counts (ρ=0.96, CCC=0.88, 95%CI: 0.8–0.93) improved substantially.

#### 3.1.5. *Semi-supervised learning*

Last, we applied a semi-supervised learning technique to improve model performance. The spacing agnostic nnU-Net generated PVS predictions or 'pseudo-labels' for images (n=12,740) from 18 datasets. The pseudo-label trained model yielded 5FCV DSC=85.6±1.4%. The pseudo-label trained model yielded the highest performance for BG-PVS (DSC=82.6±2.4%) (Table 1, Figure 8). The overall mean DSC achieved for PVS segmentation in the 7T MRI dataset A was 86.2±2.9%. The mean DSC achieved across the 3T datasets B and C was 89±2.9%. Mean DSC in datasets B and C were 89.3±2.2% and 87.7±2.5%, respectively. There was substantial agreement between predicted and actual quantities of both PVS voxel counts (ρ=0.98, CCC=0.98, 95%CI: 0.96–0.99) and cluster counts (ρ=0.96, CCC=0.89, 95%CI: 0.82–0.93).

Supplementary Figures 2-3 show Bland-Altman plots demonstrating the agreement between predicted and actual counts of PVS voxels and clusters in the white matter and basal ganglia. Supplementary Table 17 presents the corresponding correlation and concordance results.

When the final model and preprocessing pipeline was applied to raw T1w images and compared to the first set of ground truth segmentations the mean DSC was 66.2±7.0% (SEN=56.6±9.5%, PPV=81.5±6.6%). Consistent with previous models, segmentation performance was higher for WM-PVS (DSC=66.6±7.6%, SEN=57.2±11.0%, PPV=81.9±6.2%) compared to BG-PVS (DSC=58.8±12.9%, SEN=49.5±11.5%, PPV=74.7±17.6%). Overall segmentation performance was the highest in dataset B (DSC=69.4±5.6%, SEN=59.0±7.4%, PPV=84.8±3.7%), followed by dataset A (DSC=65.7±6.3%, SEN=59.8±9.7%, PPV=73.9±4.3%) and dataset C (DSC=63.5±8.1%, SEN=50.9±9.4%, PPV=85.9±3.2%). Predicted PVS voxel ($\rho=0.9$, $p<0.001$) and cluster counts ($\rho=0.91$, $p<0.001$) were strongly correlated to the manually labelled quantities in the first set of segmentations. There was strong agreement in PVS cluster counts between predicted and actual values (CCC=0.86, 95%CI: 0.74–0.93), although agreement for PVS voxel counts was moderate (CCC=0.68, 95%CI: 0.51–0.80).

**Table 1.** 5FCV DSC scores of T1w models targeting the WM and basal ganglia-PVS. Results are stratified by dataset across columns. Values are presented as Mean±SD.

| Model | Overall | A | B | C |
| --- | --- | --- | --- | --- |
| *Spacing optimisation* | | | | |
| 1.00×1.00×1.00 mm | 40.5±2.8 | 27.4±5.7 | 32.8±4.6 | 67.9±6.3 |
| 0.87×0.87×0.87 mm | 49.2±1.9 | 35.7±6.0 | 70.3±4.4 | 48.2±5.7 |
| 0.80×0.80×0.80 mm | 46.2±3.1 | 43.6±7.2 | 48.3±5.6 | 54.4±6.7 |
| 0.75×0.75×0.75 mm | 55.0±3.0 | 67.1±6.2 | 53.2±5.5 | 57.2±6.4 |
| Agnostic | 64.3±3.3 | 66.6±6.1 | 71.2±4.1 | 68.7±6.2 |
| *Preprocessing optimisation* | | | | |
| NLMF | 63.6±3.2 | 66.1±6.2 | 70.8±4.7 | 66.8±7.2 |
| AHE | 63.4±3.1 | 64.3±6.2 | 70.9±4.3 | 68.1±6.3 |
| NLMF+AHE | 63.0±3.1 | 64.9±5.9 | 70.2±5.0 | 66.9±6.7 |
| *Label cleaning* | | | | |
| Iteration 1 | 80.0±0.9 | 80.7±3.5 | 80.3±4.1 | 81.0±2.1 |
| Iteration 2 | 85.7±1.2 | 86.8±3.3 | 89.5±2.8 | 89.2±1.9 |
| *Semi-supervised learning* | | | | |
| Pseudo-labels | 85.6±1.4 | 86.2±2.9 | 89.3±2.2 | 87.7±2.5 |

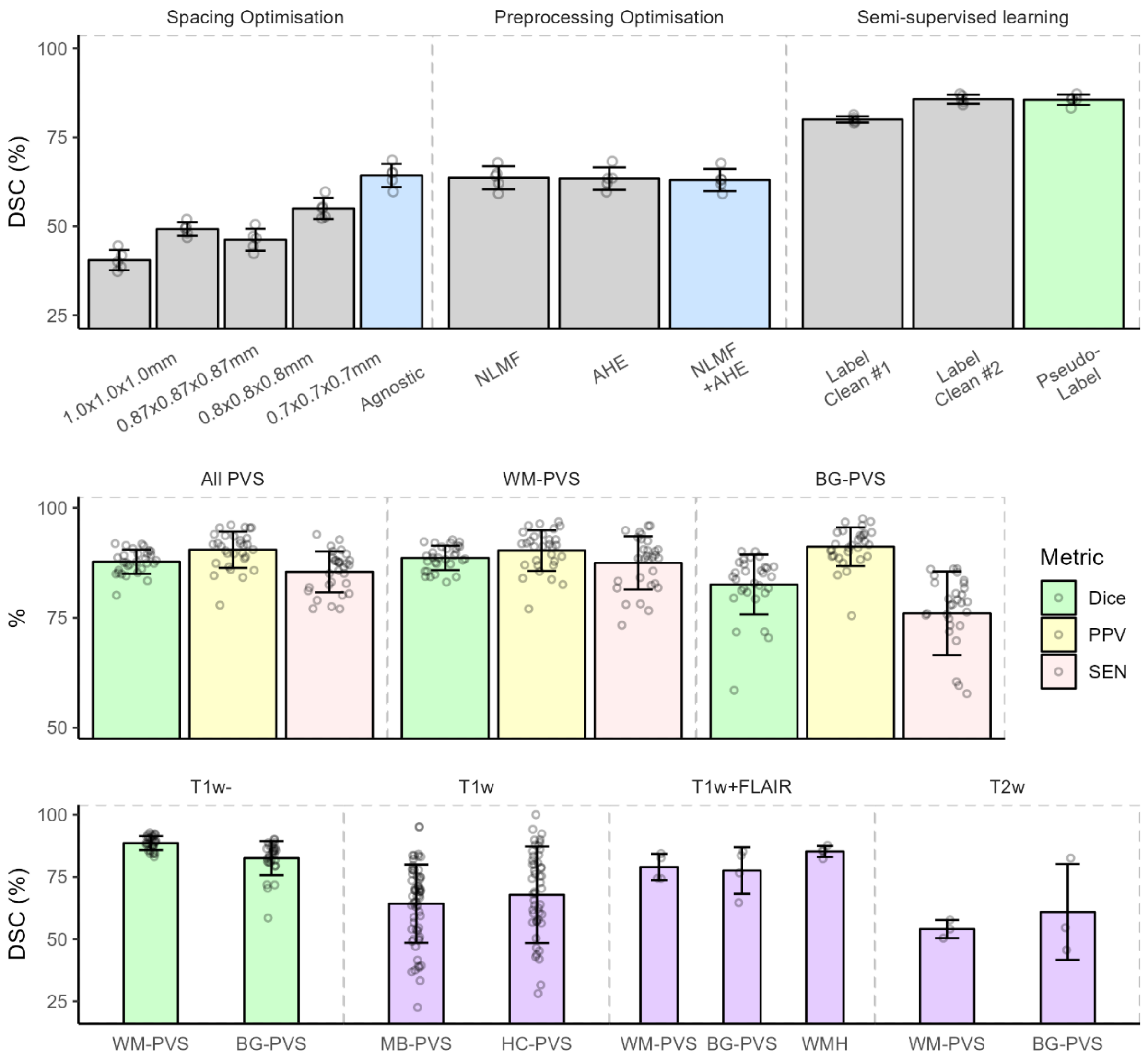

**Figure 8.** Performance of all nnU-Net models. Top: 5FCV mean Dice similarity coefficients for T1w nnU-Net models targeting the WM and BG-PVS. Middle: 5FCV metrics of the pseudo-label trained T1w nnU-Net model. Performance is shown for the overall performance (middle-left), WM-PVS (middle), and BG-PVS (middle-right). Bottom: region-wise DSC performance for the pseudo-label model (bottom-left), MB-PVS and HP-PVS models (middle-left), T1w+FLAIR model (middle-right), and T2w model (right).

### 3.2. T1w nnU-Net: MB and HP-PVS

#### 3.2.1. *Midbrain-PVS segmentation*

The median±IQR midbrain-PVS voxel and cluster counts were 53.5±42.3 and 4±2, respectively. Mean 5FCV DSC score for the MB-PVS nnU-Net was 64.3±6.5% (Table 2, Figure 8). There was poor agreement between predicted and actual quantities of MB-PVS voxel ($\rho$=0.39, CCC=0.19, 95%CI: 0.04–0.33) and cluster counts ($\rho$=0.33, CCC=0.29, 95%CI: 0.01–0.53). Supplementary Figure 4 shows Bland-Altman plots demonstrating the agreement between predicted and actual counts of MB-PVS voxels and clusters.

#### 3.2.2. *Hippocampal-PVS segmentation*

The median±IQR hippocampal-PVS voxels and cluster counts were 37.5±39.5 and 4±3, respectively. Mean 5FCV Dice score for the HP-PVS nnU-Net was 67.8±5% (Table 2, Figure 8). There was poor agreement between predicted and actual quantities of HP-PVS voxel ($\rho$=0.29, CCC=0.11, 95%CI: -0.02–0.23) and cluster counts ($\rho$=0.66, CCC=0.63, 95%CI: 0.36–0.80). Supplementary Figure 4 shows Bland-Altman plots demonstrating the agreement between predicted and actual counts of HP-PVS voxels and clusters.

### 3.3. T1w+FLAIR nnU-Net

The median±IQR number of WM-PVS and basal ganglia-PVS voxels present in the held-out images (n=4) was 4608.5±1189.5 and 802±151.25, respectively. The median number of WMH voxels was 1592±495.25. When tested on the held-out dataset, the overall mean DSC score was 81.1±4.3%. Mean DSC was highest for the WMH (DSC=85.2±2.2%), followed by BG-PVS (77.5±9.4%), and WM-PVS (79±5.3%) (Table 2, Figure 8).

### 3.4. T2w nnU-Net

The median±IQR number of WM-PVS and basal ganglia-PVS voxels present in the held-out images (n=3) was 4974±309 and 635±337.5, respectively. Visual inspections of the co-registered PVS labels revealed there were minor displacements in the images that resulted in background voxels included in the PVS labels. This was especially detrimental for PVS in the white matter. When tested on the held-out dataset, the overall mean DSC score was 54.4±4.9%. Mean DSC score for the WM-PVS and BG-PVS was 54.1±3.7% and 60.9±19.3%, respectively (Table 2, Figure 8).

**Table 2.** Performance metrics for the final nnU-Net models. Metrics for the T1w models were derived from 5FCV. Metrics for the T1w+FLAIR and T2w models were derived from a train/test procedure. DSC=Dice-Similarity coefficient. SEN=Sensitivity. PPV=positive predictive value. Values are presented as Mean±SD.

| Modality/ROI | DSC | SEN | PPV |
|---|---|---|---|
| *T1w* | | | |
| Overall | 85.6±1.4 | 86.0±1.1 | 89.7±2.5 |
| WM-PVS | 88.6±0.9 | 88.5±1.9 | 89.4±2.8 |
| BG-PVS | 82.6±2.4 | 74.5±4.2 | 91.1±2.0 |
| *T1w* | | | |
| MB-PVS | 64.3±6.5 | 53.6±8.6 | 81.1±2.7 |
| *T1w* | | | |
| HP-PVS | 67.8±5.0 | 57.6±6.4 | 87.8±5.9 |
| *T1w+FLAIR* | | | |
| Overall | 81.1±4.3 | 74.0±6.2 | 90.0±1.8 |
| WM-PVS | 79.0±5.3 | 71.2±7.6 | 89.0±2.6 |
| BG-PVS | 77.5±9.4 | 68.1±13.6 | 91.8±1.3 |
| WMH | 85.2±2.2 | 80.3±3.7 | 90.9±2.0 |
| *T2w* | | | |
| Overall | 54.4±4.9 | 43.0±4.9 | 74.4±3.9 |
| WM-PVS | 54.1±3.7 | 42.6±3.5 | 73.9±3.2 |
| BG-PVS | 60.9±19.3 | 50.2±20.5 | 79.3±13.3 |

## 4. Discussion

By combining the nnU-Net residual encoder with image-processing techniques, we developed a comprehensive pipeline for the automated segmentation of perivascular spaces in MRI. We showed that a sparse annotation strategy can efficiently acquire full brain PVS segmentations from partially complete labels. Second, voxel-spacing is a crucial parameter in the image preparation phase. Resampling images to a non-native resolution proved detrimental to model performance and PVS segmentation. Our voxel-spacing agnostic nnU-Net model, which avoids image resampling, yields the most robust performance across heterogenous MRI parameters. Thirdly, we integrated image enhancement techniques and a semi-supervised training paradigm to boost model performance. Lastly, we developed pilot models capable of PVS segmentation in the midbrain and hippocampal regions in T1w MRI, as well as models based on both T1w and FLAIR scans, or a single T2w MRI scan. Our models represent a highly versatile research toolkit capable of comprehensive examination of perivascular spaces using multiple MRI sequences.

## 4.1. T1w nnU-Net: WM+BG-PVS

### *4.1.1. Sparse annotation strategy*

A major challenge in medical image segmentation tasks is the acquisition of high-quality segmentation labels. Manual data labelling is laborious and time-consuming. Previous MRI-PVS segmentation algorithms relied on 40 training images or less (Waymont et al., 2024). To avoid the need for manual segmentations, some have proposed models validated on synthetic MRI scans with simulated PVS structures (Bernal et al., 2022; Park et al., 2016; Zhang et al., 2017). Here, we addressed the manually intensive process of PVS segmentation via a sparse annotation strategy.

Sparse annotation strategies are a recent innovation that enable the efficient acquisition of labelled training data. Gotkowski et al., (2024) adapted the nnU-Net to complete multi-organ segmentations based on manually annotated scribbles outlining abdominal organs in computed tomography scans (Gotkowski et al., 2024). We are the first to implement a sparse annotation strategy for the acquisition of MRI-PVS labels. We demonstrate that the sparse annotation strategy is effective in obtaining high-quality PVS segmentations. Future research requiring the procurement of additional PVS segmentations can be expedited with a sparse annotation strategy.

### *4.1.2. Target spacing optimisation*

Target voxel spacing is a crucial parameter in the image preparation phase. MRI-PVS segmentation models have typically been developed on a single homogenous image dataset. Previous deep learning PVS segmentation approaches resampled input MRI scans to a pre-selected voxel spacing for example 1.0 mm isotropic voxel spacing (Boutinaud et al., 2021). Whilst the nnU-Net, by default, selects the median voxel spacing of the training dataset as the target resolution. However, this approach leads to resampling artefacts and likely lessens segmentation performance.

To determine the most robust and generalisable voxel spacing, we benchmarked the nnU-Net models across several target voxel spacings (0.7, 0.8, 0.87, 1.0 mm isotropic and an agnostic variant). Of all variants, the voxel spacing agnostic nnU-Net variant offered the most generalisable PVS segmentation performance (Figure 8). Unlike models which considered voxel sizes in the image preprocessing pipeline, the spatially agnostic model did not suffer from resampling artefacts. Researchers developing machine learning models for the automated segmentation of PVS in MRI can benefit from applying a voxel-size agnostic approach to facilitate model generalisability. This also allows the pipeline to be applied to MRI images acquired at various field strengths.

#### *4.1.3. Image preprocessing*

Several image preprocessing techniques have been applied to improve PVS visibility in MRI. For example, gamma contrast adjustment has been employed to improve the appearance of PVS, thus facilitating manual grading (Wang et al., 2016). Sepehrband et al., (2019) applied non-local means filtering (NLMF) for image denoising, followed by combining co-registered T1w and T2w scans to enhance the appearance of PVS (Sepehrband et al., 2019). Whereas contrast-limited adaptive histogram equalisation has previously been applied to T2w MRI to improve the visibility of PVS and facilitate their segmentation with a support vector machine classifier (González-Castro et al., 2016).

Similarly, we integrated image enhancement techniques to improve the automated detection and annotation of MRI-PVS. We applied NLMF to reduce the noise surrounding PVS, followed by AHE to emphasise its appearance with respect to non-PVS structures. Compared to the model trained with raw T1w images, no quantitative benefit was observed for models trained with pre-processed images (Table 1). However, image enhancement greatly assisted in manual label cleaning by reducing the partial volume effects. The boundary between PVS structures and surrounding white matter was clearer in images pre-processed with NLMF and AHE (Figure 3). Thus, the enhanced images were used to assist manual label cleaning.

#### *4.1.4. Iterative label cleaning*

The development of reliable machine learning models hinges on the use of high-quality training data. Noisy labels are a major problem in the biomedical imaging field (Karimi et al., 2020). The presence of perivascular spaces in MRI are often obscured by other imaging lesions or mimics.

We demonstrate model performance increases substantially after each round of iterative label cleaning. After two rounds of label cleaning, our T1w nnU-Net achieved an overall DSC score of 85.7±1.2% (Table 1). This is higher than all previously reported results, including those using high quality 7T, T2w MRI (Pham et al., 2022; Waymont et al., 2024). Future research should include at least one stage of iterative label cleaning when developing reliable deep learning models for automated segmentation tasks, especially when training data are limited. Additionally, researchers could focus on pre-processed images, instead of raw MRI scans, when acquiring a gold standard set of segmentations to minimise the impact of noisy labels.

#### *4.1.5. Semi-supervised learning*

Our final technique to optimise the T1-weighted MRI nnU-Net model was based on a semi-supervised training paradigm which included pseudo-labels in the training dataset. Pseudo-labelling is a recently proposed technique in computational image segmentation tasks (Ferreira et al., 2023). The goal of pseudo-labelling is to boost model performance without the need for additional manually labelled training data. We supplemented our original 30 ground truth segmentations with pseudo-labels (n=12,740) acquired from 18 datasets. Although, no improvement to the Dice Similarity Coefficient was observed, retraining the nnU-Net with pseudo-labels benefited the concordance between predicted and manually segmented measurements of PVS voxel and cluster counts. The semi-supervised training approach likely reinforced the model's latent representation of PVS structures. Notably, our large and diverse training dataset presents an important advance in the development of machine learning models with the purpose of automating PVS segmentation in MRI.

Our final evaluation of the T1w nnU-Net compared the model's predictions to the original ground truth segmentations derived from raw T1w images. Combining image preprocessing, iterative label cleaning, and pseudo labelling resulted in substantially improved segmentation performance (from mean DSC of 64.3% to 85.6%). However, when evaluating the final model against the 30 initial ground truth PVS segmentations, the mean DSC score was 66.2%.

Our findings highlight that not all PVS are equally visible. Some voxels are detected more reliably than others. The observed improvement in model performance through iterative label cleaning likely arises by removing poor-quality or noisy labels rather than increasing the model's sensitivity to PVS. Consequently, the enhanced segmentation metrics reflect improved reproducibility and PVS segmentation quality as opposed to an improvement in the model's ability to detect PVS.

### 4.2. T1w nnU-Net: MB and HP-PVS

To date, the study of midbrain and hippocampal PVS has relied on cluster counts or binarised scores indicating their presence or absence (Adams et al., 2013; Dubost et al., 2019). Enlargement of perivascular spaces in the midbrain was found to be significantly larger in Parkinson's disease patients compared to age and sex matched controls (Shen et al., 2021). By contrast, enlargement of the perivascular spaces in the hippocampus have been associated with older age and hypertension status (Evans et al., 2022; Yao et al., 2014). Although, the relationship between HP-PVS and cognition, mild cognitive impairment, and vessel pulsatility index measured by transcranial doppler imaging are inconsistent, with reports of no significant associations (Jiménez-Balado et al., 2018; Sim et al., 2020). These studies are limited by the coarse measures of PVS enlargement. Our novel nnU-Net models provide voxel-wise quantifications of PVS in both the midbrain and hippocampal regions. Thus, enabling highly detailed examination of midbrain and hippocampal PVS in the context of disease.

### 4.3. T1w+FLAIR nnU-Net

Rashid et al., (2023) implemented a multi-modal convolutional neural network with T1w and FLAIR for PVS segmentation (Rashid et al., 2023). Aside from differences in neural network architecture, we provide two key advantages. First, we implemented region-based labelling for PVS in the WM and basal ganglia. Unique WM and BG-PVS labels enable quantification of regional PVS metrics without the need for brain tissue parcellation algorithms, e.g. FreeSurfer. Second, our T1w+FLAIR model can segment WMHs alongside the PVS labels. Thus, supplementary modules dedicated to WMH detection via FLAIR sequences, e.g., Lesion Segmentation Tool, are not required to remove the confounding presence of WMH. Our dual-channel nnU-Net incorporates T1w and FLAIR sequences to distinguish PVS from WMH. This model therefore mitigates the confounding presence of white matter lesions on PVS detection. The resulting segmentation maps allow for direct assessments of the topological relationships between these two markers of cerebral small vessel disease.

### 4.4. T2w nnU-Net

PVSs are significantly more prominent and easily visualised on T2w rather than T1w. Despite the greater prevalence of T1w scans, previous models of PVS segmentation in MRI have largely focussed on the T2w sequence (Cai et al., 2015; Duering et al., 2023; Lian et al., 2018; Wardlaw et al., 2013; Zong et al., 2016). We extended out nnU-Net to T2w images using MRI scans registered to the original PVS segmentations from the T1w MRI data. However, a significant issue in the T2w training data relates to the effect of co-registration errors. Slight misalignments in the multi-modal MRI co-registrations resulted in cumulative and substantial errors in PVS labels, consequently eroding the overall training data quality and model performance. Future advances in image co-registration may address this issue and allow the effective use of T2w with T1w images.

### 4.5. Limitations

One limitation relates to the small sample size used to validate model performance. Rigorous and robust machine learning validation is dependent on the availability of datasets of sufficient size and diversity (Isensee et al., 2024). Our T1w model targeting WM and BG-PVS was validated with only 30 segmentations available. Nevertheless, our model performed remarkably well despite the relatively low number of segmentations available. Moreover, our validation dataset comprised images from healthy controls rather than patients with neurological disorders. Future studies could include MRI from disease populations. This would likely improve the model's ability to handle MRI scans encompassing pathological lesions alongside PVS structures.

The model targeting midbrain and hippocampal PVS can be improved in several ways. The first being the inclusion of additional labelled training data. In the field of biomedical image segmentation, the training dataset size (n=50-60) used here is relatively small. Second, label cleaning can be used to improve the quality of the segmentation labels and resulting model performance. We implemented label cleaning for the T1w model targeting PVS in the white matter and basal ganglia only. Lastly, image preprocessing techniques, such as NLMF and AHE, can be explored to enhanced PVS detection in these brain regions. Whether these techniques would benefit automated midbrain and hippocampal-PVS segmentation is unknown.

A limitation of our T1w+FLAIR and T2w models is the limited number of test samples. A 5-fold cross-validation offers a more robust and reliable evaluation of model performance. However, it was not possible in this case due to the insufficient number of ground truth segmentations. Instead, we relied on a train-test procedure to evaluate performance. Future work could acquire additional labels by a sparse annotation strategy, enabling a more rigorous k-fold cross validation procedure. Alternatively, techniques such as label cleaning of the available segmentations and pseudo-labels could also be explored to generate higher-quality training data.

## 5. Conclusion

Perivascular spaces in MRI are an emerging imaging biomarker for several neurological conditions. We demonstrated a novel method for the efficient acquisition of high-quality PVS labels via a sparse segmentation strategy. Iterative label cleaning was found to be highly effective in improving nnU-Net segmentation of white matter and basal ganglia PVS in T1w

MRI. Our final model is capable of robust PVS segmentation across heterogenous imaging protocols. Further, we developed pilot nnU-Net models for the automated segmentation of PVS in the midbrain and hippocampal regions in T1-weighted MRI. Our models have been made freely available to researcher enabling comprehensive and rigorous research of perivascular spaces in MRI.

## Data Availability Statement

Our nnU-Net models are open-source and freely available for research use at our GitHub repository: https://github.com/wpham17/nnUNet-Perivascular-Spaces.

**Acknowledgements**

**A4**

The A4 Study is a secondary prevention trial in preclinical Alzheimer's disease, aiming to slow cognitive decline associated with brain amyloid accumulation in clinically normal older individuals. The A4 Study is funded by a public-private-philanthropic partnership, including funding from the National Institutes of Health-National Institute on Aging, Eli Lilly and Company, Alzheimer's Association, Accelerating Medicines Partnership, GHR Foundation, an anonymous foundation and additional private donors, with in-kind support from Avid and Cogstate. The companion observational Longitudinal Evaluation of Amyloid Risk and Neurodegeneration (LEARN) Study is funded by the Alzheimer's Association and GHR Foundation. The A4 and LEARN Studies are led by Dr. Reisa Sperling at Brigham and Women's Hospital, Harvard Medical School and Dr. Paul Aisen at the Alzheimer's Therapeutic Research Institute (ATRI), University of Southern California. The A4 and LEARN Studies are coordinated by ATRI at the University of Southern California, and the data are made available through the Laboratory for Neuro Imaging at the University of Southern California. The participants screening for the A4 Study provided permission to share their de-identified data in order to advance the quest to find a successful treatment for Alzheimer's disease. We would like to acknowledge the dedication of all the participants, the site personnel, and all of the partnership team members who continue to make the A4 and LEARN Studies possible. The complete A4 Study Team list is available on: www.actcinfo.org/a4-study-team-lists.

**ABIDE I**

ABIDE I Funding Acknowledgements: Primary support for the work by Adriana Di Martino was provided by the (NIMH K23MH087770) and the Leon Levy Foundation. Primary support for the work by Michael P. Milham and the INDI team was provided by gifts from Joseph P. Healy and the Stavros Niarchos Foundation to the Child Mind Institute, as well as by a NIMH award to MPM (NIMH R03MH096321).

**ABVIB**

Data collection and sharing for the ABVIB project was funded by the National Institutes on Aging (NIA) P01 AG12435.

ADNI

**ADNI**

Data collection and sharing for the Alzheimer's Disease Neuroimaging Initiative (ADNI) is funded by the National Institute on Aging (National Institutes of Health Grant U19AG024904). The grantee organization is the Northern California Institute for Research and Education. In the past, ADNI has also received funding from the National Institute of Biomedical Imaging and Bioengineering, the Canadian Institutes of Health Research, and private sector contributions through the Foundation for the National Institutes of Health (FNIH) including generous contributions from the following: AbbVie, Alzheimer's Association; Alzheimer's Drug Discovery Foundation; Araclon Biotech; BioClinica, Inc.; Biogen; Bristol-Myers Squibb Company; CereSpir, Inc.; Cogstate; Eisai Inc.; Elan Pharmaceuticals, Inc.; Eli Lilly and Company; EuroImmun; F. Hoffmann-La Roche Ltd and its affiliated company Genentech, Inc.; Fujirebio; GE Healthcare; IXICO Ltd.; Janssen Alzheimer Immunotherapy Research & Development, LLC.; Johnson & Johnson Pharmaceutical Research & Development LLC.; Lumosity; Lundbeck; Merck & Co., Inc.; Meso Scale Diagnostics, LLC.; NeuroRx Research; Neurotrack Technologies; Novartis Pharmaceuticals Corporation; Pfizer Inc.; Piramal Imaging; Servier; Takeda Pharmaceutical Company; and Transition Therapeutics.

**ADNI/ADNIDOD**

Data collection and sharing for this project was funded by the Alzheimer's Disease Neuroimaging Initiative (ADNI) (National Institutes of Health Grant U01 AG024904) and DOD ADNI (Department of Defense award number W81XWH-12-2-0012). ADNI is funded by the National Institute on Aging, the National Institute of Biomedical Imaging and Bioengineering, and through generous contributions from the following: Alzheimer's Association; Alzheimer's Drug Discovery Foundation; BioClinica, Inc.; Biogen Idec Inc.; Bristol-Myers Squibb Company; Eisai Inc.; Elan Pharmaceuticals, Inc.; Eli Lilly and Company; F. Hoffmann-La Roche Ltd and its affiliated company Genentech, Inc.; GE Healthcare; Innogenetics, N.V.; IXICO Ltd.; Janssen Alzheimer Immunotherapy Research & Development, LLC.; Johnson & Johnson Pharmaceutical Research & Development LLC.; Medpace, Inc.; Merck & Co., Inc.; Meso Scale Diagnostics, LLC.; NeuroRx Research; Novartis Pharmaceuticals Corporation; Pfizer Inc.; Piramal Imaging; Servier; Synarc Inc.; and Takeda Pharmaceutical Company. The Canadian Institutes of Health Research is providing funds to support ADNI clinical sites in Canada. Private sector contributions are facilitated by the Foundation for the National Institutes of Health (www.fnih.org). The grantee organization is the Northern California Institute for Research and Education, and the study is coordinated by the Alzheimer's Disease Cooperative Study at the University of California, San Diego. ADNI data are disseminated by the Laboratory for Neuro Imaging at the University of Southern California. This research was also supported by NIH grants P30 AG010129 and K01 AG030514.

**AIBL**

Data used in the preparation of this article was obtained from the Australian Imaging Biomarkers and Lifestyle flagship study of ageing (AIBL) funded by the Commonwealth Scientific and Industrial Research Organisation (CSIRO) which was made available at the ADNI database (www.loni.usc.edu/ADNI). The AIBL researchers contributed data but did not participate in analysis or writing of this report. AIBL researchers are listed at www.aibl.csiro.au.

**HCP1200**

Data were provided in part by the Human Connectome Project, WU-Minn Consortium (Principal Investigators: David Van Essen and Kamil Ugurbil; 1U54MH091657) funded by the 16 NIH Institutes and Centers that support the NIH Blueprint for Neuroscience Research; and by the McDonnell Center for Systems Neuroscience at Washington University.

**NIFD**

Data collection and sharing for the NIFD project was funded by the Frontotemporal Lobar Degeneration Neuroimaging Initiative (National Institutes of Health Grant R01 AG032306). The study is coordinated through the University of California, San Francisco, Memory and Aging Center. FTLDNI data are disseminated by the Laboratory for Neuro Imaging at the University of Southern California.

**OASIS-3**

Data were provided in part by OASIS-3: Longitudinal Multimodal Neuroimaging: Principal Investigators: T. Benzinger, D. Marcus, J. Morris; NIH P30 AG066444, P50 AG00561, P30 NS09857781, P01 AG026276, P01 AG003991, R01 AG043434, UL1 TR000448, R01 EB009352. AV-45 doses were provided by Avid Radiopharmaceuticals, a wholly owned subsidiary of Eli Lilly.

**PPMI**

Data used in the preparation of this article were obtained on June, 7th 2024 from the Parkinson's Progression Markers Initiative (PPMI) database (www.ppmi-info.org/access-dataspecimens/download-data), RRID:SCR 006431. For up-to-date information on the study, visit www.ppmi-info.org "PPMI – a public-private partnership – is funded by the Michael J. Fox Foundation for Parkinson's Research and funding partners, including 4D Pharma, Abbvie, AcureX, Allergan, Amathus Therapeutics, Aligning Science Across Parkinson's, AskBio, Avid Radiopharmaceuticals, BIAL, BioArctic, Biogen, Biohaven, BioLegend, BlueRock Therapeutics, Bristol-Myers Squibb, Calico Labs, Capsida Biotherapeutics, Celgene, Cerevel Therapeutics, Coave Therapeutics, DaCapo Brainscience, Denali, Edmond J. Safra Foundation, Eli Lilly, Gain Therapeutics, GE HealthCare, Genentech, GSK, Golub Capital, Handl

Therapeutics, Insitro, Jazz Pharmaceuticals, Johnson & Johnson Innovative Medicine, Lundbeck, Merck, Meso Scale Discovery, Mission Therapeutics, Neurocrine Biosciences, Neuron23, Neuropore, Pfizer, Piramal, Prevail Therapeutics, Roche, Sanofi, Servier, Sun Pharma Advanced Research Company, Takeda, Teva, UCB, Vanqua Bio, Verily, Voyager Therapeutics, the Weston Family Foundation and Yumanity Therapeutics.